\documentclass[aps,prl,twocolumn,nopacs,superscriptaddress,groupedaddress]{revtex4}  
\usepackage{graphicx}  
\usepackage{dcolumn}   
\usepackage{bm}        
\usepackage{amssymb}   
\usepackage{epstopdf}
\usepackage{color}
\usepackage{dcolumn}
\usepackage{bm}
\usepackage{amsmath,braket,xcolor,subfigure,upgreek}

\def\gapx{\lower 2pt \hbox{$\buildrel>\over{\scriptstyle{\sim}}$\ }}
\def\lapx{\lower 2pt \hbox{$\buildrel<\over{\scriptstyle{\sim}}$\ }}

\RequirePackage[colorlinks,linkcolor=black,citecolor=blue,hyperindex]{hyperref}

\begin{document}
\title{Conductivity in Organic Semiconductors Hybridized with the Vacuum Field}

\author{E. Orgiu$^\dagger$} 
\affiliation{ISIS \& icFRC, Universit\'e de Strasbourg and CNRS, 67000 Strasbourg, France}
\author{J. George$^\dagger$}%
\affiliation{ISIS \& icFRC, Universit\'e de Strasbourg and CNRS, 67000 Strasbourg, France}
\author{J. A. Hutchison$^\dagger$}%
\affiliation{ISIS \& icFRC, Universit\'e de Strasbourg and CNRS, 67000 Strasbourg, France}
\author{E. Devaux}%
\affiliation{ISIS \& icFRC, Universit\'e de Strasbourg and CNRS, 67000 Strasbourg, France}
\author{J. F. Dayen}%
\affiliation{IPCMS \& icFRC, Universit\'e de Strasbourg and CNRS, 67034 Strasbourg, France}%
\author{B. Doudin}%
\affiliation{IPCMS \& icFRC, Universit\'e de Strasbourg and CNRS, 67034 Strasbourg, France}%
\author{F. Stellacci}%
\affiliation{EPFL, STI SMX-GE MXG 030 Station 12, CH-1015 Lausanne, Switzerland}%
\author{C. Genet}%
\affiliation{ISIS \& icFRC, Universit\'e de Strasbourg and CNRS, 67000 Strasbourg, France}
\author{J. Schachenmayer}%
\affiliation{JILA, NIST, Department of Physics, University of Colorado, 440 UCB, Boulder, CO 80309, USA}%
\author{C. Genes}%
\affiliation{Institut f\"ur Theoretische Physik, Universit\"at Innsbruck, Technikerstrasse 25, A-6020 Innsbruck, Austria}%
\author{G. Pupillo}%
\affiliation{ISIS \& icFRC, Universit\'e de Strasbourg and CNRS, 67000 Strasbourg, France}
\affiliation{IPCMS \& icFRC, Universit\'e de Strasbourg and CNRS, 67034 Strasbourg, France}%
\author{P. Samor\`i}%
\affiliation{ISIS \& icFRC, Universit\'e de Strasbourg and CNRS, 67000 Strasbourg, France}
\author{T. W. Ebbesen$^*$}%
\affiliation{ISIS \& icFRC, Universit\'e de Strasbourg and CNRS, 67000 Strasbourg, France}

\date{\today}

\begin{abstract}
Organic semiconductors have generated considerable interest for their potential for creating inexpensive and flexible devices easily processed on a large scale~\cite{1,2,3,4,5,6,7,8,26,27,28}. However technological applications are currently limited by the low mobility of the charge carriers associated with the disorder in these materials~\cite{5,6,7,8}. Much effort over the past decades has therefore been focused on optimizing the organisation of the material or the devices to improve carrier mobility. Here we take a radically different path to solving this problem, namely by injecting carriers into states that are hybridized to the vacuum electromagnetic field. These are coherent states that can extend over as many as $10^5$ molecules and should thereby favour conductivity in such materials. To test this idea, organic semiconductors were strongly coupled to the vacuum electromagnetic field on plasmonic structures to form polaritonic states with large Rabi splittings $\sim$ 0.7 eV. Conductivity experiments show that indeed the current does increase by an order of magnitude at resonance in the coupled state, reflecting mostly a change in field-effect mobility as revealed when the structure is gated in a transistor configuration. A theoretical quantum model is presented that confirms the delocalization of the wave-functions of the hybridized states and the consequences on the conductivity.  While this is a proof-of-principle study, in practice conductivity mediated by light-matter hybridized states is easy to implement and we therefore expect that it will be used to improve organic devices. More broadly our findings illustrate the potential of engineering the vacuum electromagnetic environment to modify and to improve properties of materials. 
\end{abstract}


\maketitle

Light and matter can enter into the strong coupling regime by exchanging photons faster than any competing dissipation processes. This is normally achieved by placing the material in a confined electromagnetic environment, such as a Fabry-Perot (FP) cavity composed of two parallel mirrors that is resonant with an electronic transition in the material. Alternatively, one can use surface plasmon resonances as in this study. Strong coupling leads to the formation of two hybridized light-matter polaritonic states, P+ and P-, separated by the so-called Rabi splitting, as illustrated in Figure 1a. According to quantum electrodynamics, in the absence of dissipation, the Rabi splitting for a single molecule is given by
\begin{equation}
\hbar \Omega_{\rm R}=2 \sqrt{\frac{\hbar\omega}{2 \epsilon_0 v}} \cdot d \cdot \sqrt{n_{\rm ph}+1}
\end{equation}
where $\hbar \omega$ is the cavity resonance or transition energy ($\hbar$ the reduced Planck constant), $\epsilon_0$ the vacuum permittivity, $v$ the mode volume, $d$ the transition dipole moment of the material and $n_{\rm ph}$ the number of photons present in the system. The last term implies that, even in the dark, the Rabi splitting  has a finite value which is due to the interaction with the vacuum electromagnetic field. This vacuum Rabi splitting can be further increased by coupling a large number $N$ of oscillators to the electromagnetic mode since $\hbar \Omega_{\rm R}^N \propto \sqrt{N}$~\cite{9}.  In this ensemble coupling, in addition to P+ and P-, there are many other combinations of states, known as dark states \cite{HoudrePSS2005}, that are located in the middle of the Rabi splitting as illustrated in Fig.~1a. 

Polaritons have been extensively studied in inorganic and organic materials~\cite{11,12,13,14,15,16,17,18,19,20,21,22,23,24,25,31,32,33,34,35,36,Tischler,Torma}. Vacuum Rabi splittings as large as 1 eV have been reported for strongly coupled molecules~\cite{14,15}, thereby significantly modifying the electronic structure of the molecular material as can be seen in the work function, the ground state energy shift as well as in the chemical reactivity~\cite{16,17,18}.  In such situations, P+ and P- are coherent collective states (Fig.~1b) that extend over the mode volume and may involve $\sim 10^5$ oscillators as in the present systems. This collective nature has been demonstrated by the coherent fluorescence of P- over micrometers distances in strongly coupled molecular states~\cite{19,Torma}.  Charge carrier mobility should also benefit from the extended coherence associated with
light-hybridized states as compared to the normal carrier mobilities limited by the hopping between molecules and by scattering induced by molecular disorder. 

While electron injected polariton LEDs and lasers have been reported in the last years~\cite{31,32,33,34,35,36,37}, conductivity mediated by light-hybridized materials has not been considered.  One of the challenges in those studies has been to inject electrons into P+ or P- since these states have not just well defined energies but also momenta associated with their dispersion curves. In this regard, molecular materials have the advantage that the energies of the states are inhomogenously broadened and fluctuating in time resulting in spectral widths that facilitates polariton creation via electron recombination. Indeed the very first study of polariton electroluminescence was carried out on organic materials~\cite{Tischler}.

In this study, an initial screening was carried out on different organic semiconductors known to be air-stable and display good field-effect mobilities in solution processed devices ($10^{-2} \sim 10^{-1}$ ${\rm cm}^2$V$^{-1}$s$^{-1}$ for optimized field effect geometries). We focus here on three members of the aromatic di-imide family, an example of which,  PDI2EH-CN$_2$, is shown in Fig. 1c~\cite{26,27,28}. The synthesis of this molecule is described in Supplementary Section I. The other molecules are the brominated version of the former, PDI2EH-BR$_2$, and the di-imide naphthalene polymer, P(NDI2OD-T2) (ActivInk N2200 from Polyera Inc.). Noteworthy, we have chosen these organic materials in view of their very different charge transport capacities but also because they all possess a LUMO energy level at $\sim 4$ eV (see Supplementary Section IV). This last observation rules out the injection capacity as a main parameter for the interpretation of the results appearing later in the manuscript when comparing their relative response. Instead, as we will see, the differences that are important are the intrinsic charge transport capacity of each molecular system, influenced by the film disorder. These compounds possess intense absorption bands (Fig.~1d) that can be used to strongly couple them to surface plasmon (SP) resonances. For this we used a hexagonal array of holes (Fig.~1e) in either Al or Ag films which give rise to well-defined SP resonances~\cite{29} also shown in Fig.~1d. The low work-function of an Al electrode is generally favorable for electron injection in n-type semiconductors. In the case of an Ag electrode, we still have good electron injection, as confirmed in our experiments (see Supplementary Section IV, Fig. S6). Such efficient electron injection from high work function metals into PDI derivatives has been previously reported ~\cite{9,11}. 

%
%
%
%
Figures 1f and 1g show the dispersion of surface plasmon resonances of the Ag hole array as a function of hole period and the same when the array is covered with a thin film of PDI2EH-CN$_2$. Unlike with Fabry-Perot cavities, in the case of strong coupling with SPs on hole arrays the upper polariton is less visible due to Fano type interferences~\cite{14}. However, a Rabi splitting of $\sim$0.7 eV can be estimated, corresponding to 30$\%$ of the transition energy, typical of the ultra-strong coupling regime~\cite{14,15}.

In order to assess the conductivity of strongly coupled organic semiconductors, we compare it to the one of the bare molecular material, by recording current-voltage (I-V) curves for thin films of the semiconductors between two electrodes as illustrated in Fig. 2a. First, a 100 nm thick layer of Al or Ag was evaporated on a glass substrate through a shadow mask to yield a long 80 $\mu$m wide strip. Then hexagonal arrays with period $P$ were milled over a 55 $\mu$m long area on the metal strip. The active channel, 50 $\mu$m long, was defined by milling gaps g of 200 nm separating this area from the source and drain electrodes (Fig.~2a). The width of these gaps was found to be very important since a distance larger than the decay of the evanescent field ($\sim$ the wavelength) of the surface plasmon would result in the current flowing through an area where the material would not be hybridized. Notice that the array continues for a few microns on both electrodes in order to ensure that the current is already injected into the strongly coupled system before reaching the active channel (Fig.~2a).   

The I-V curves were then recorded for different array periods and compared to that of an unstructured metal film also covered with the organic semiconductors (the reference).  An example is shown for selected periods in Fig.~2b in the case of the PDI2EH-CN$_2$.  Similar experiments were carried out for the other two compounds. By plotting the current average at +/-15 V versus $2\pi / P$, $P$ the period of the array, two current resonances are clearly visible as can be seen in Fig. 2c for both PDI2EH-CN$_2$ and P(NDI2OD-T2). Interestingly, the current peaks happen for different periods. These correspond exactly to the intersections of the molecular absorption peaks and the surface plasmon modes (1,0) and (1,1) of the substrate side of the hole array where strong coupling occurs (Fig. 2d). In other words, the current is boosted when the organic semiconductor is strongly hybridized with the surface plasmon modes. The current increases by over an order of magnitude in this regime. The third compound, PDI2EH-Br$_2$, showed no detectable change in current with conductivity in the pA range (see Supplementary Fig. S5). The differences between these compounds will be discussed below. Please note that in Fig. 2c, Ag and Al were chosen as the metal electrodes because in each case, the charge injection is optimal for each electrode/semiconductor combination (see Supplementary Section IV). In addition, the injection from different metals was tested for PDI2EH-CN$_2$ and P(NDI2OD-T2), as reported in the Supplementary Fig. S2 for PDI2EH-CN$_2$ and Fig. S6 for P(NDI2OD-T2).

In the case of PDI2EH-CN$_2$, the same experiment was repeated on hexagonal hole arrays milled in an Al film with the spectra shown in Fig.~3a. The system shows (Fig.~3b) the strongest current resonance for a period of 280 nm which again corresponds to the strongly coupled condition. The current increases by roughly an order of magnitude as compared to the reference consisting of PDI2EH-CN$_2$ on a flat Al film. In order to further assess the electronic properties of the coupled system, a 3-terminal gated configuration was prepared as illustrated schematically in Fig.~3c. Figures 3d and 3e show the current versus the gating voltage for the on-resonance sample in Al and the reference, respectively. The mobility extracted from this data shows an increase from 2.0$\times10^{-3}$ cm$^2$V$^{-1}$s$^{-1}$ -a value in agreement with the values extracted in a standard geometry (see Supplementary Table I)- to 1.7$\times10^{-2}$ cm$^2$V$^{-1}$s$^{-1}$. In other words the increased carrier mobility in the coupled system is responsible for most of the enhanced current. 

The reference in the above data is always the unstructured metal film covered with the organic semiconductors. This reference shows a current larger than just the semiconductor, enhanced by the metallic substrate, but not shorted by the metal. Indeed, the measured field-effect mobility results from the intrinsic organic semiconducting properties. We have also checked random arrays of holes which show no current increases as expected since there are no well-defined plasmon resonances in such systems~\cite{Random}. 

To explain the above results, we introduced a simple theoretical model where each molecule is represented by two states forming the conduction and valence bands of the semiconductor. The molecular states are resonantly coupled by the vacuum field with the relevant Hamiltonian
\begin{eqnarray}
 H= \left( -  \sum_{j=1}^{M-1} J^e_j ( c^e_j)^\dag  c^e_{j+1} -  \sum_{j=1}^{M-1} J^h_j ( c^h_j)^\dag  c^h_{j+1} \right. \nonumber\\
\left. +  \sum_j  g_j  \; a^\dag  c^e_j  c^h_j + \text{h.c.} \right)
+ \sum_{j=1}^M (\epsilon^e_j  n^e_j + \epsilon_j^h  n^h_j).
\label{ham}
\end{eqnarray}
Here, in a second-quantized formalism, $c_j^e$($ c_j^{e\dagger}$)  and $ c_j^{h}$($c_j^{h\dagger}$) correspond to the destruction (creation) of an electron in the conduction band and of a hole in the valence band, respectively, with $n_j^e=c_j^{e\dagger}c_j^e$ ($n_j^h=c_j^{h\dagger}c_j^h$)  the local electron (hole) occupation, and $j$ the site index (i.e., the molecule location) for a system of size $M$, $M$ corresponding to the number of molecules.  The local coupling $g_j \propto d_{eh} \exp(-ik\ell j) $ (with strength $d_{eh}$) is mediated by the surface plasmon field (with a delocalized surface plasmon wave vector $k$) resonant with the interband transition~\cite{33,37}, here represented by a destruction (creation) operator $a$ ($a^\dagger$). $\ell$ is the intermolecular spacing, in the nm range, and thus $k\ell\ll 1$. The third term represents processes where a surface plasmon is coherently created when an electron is transferred from the conduction to the valence band ($a^\dagger c_j^e c_j^h$) and, vice-versa, where a surface plasmon is coherently absorbed  when an electron is promoted from the valence to the conduction band ($c_j^{e\dagger} c_j^{h\dagger}a$). The last term represents possible local shifts of the energy $\epsilon_j^e$ ($\epsilon_j^h$) for the electrons (holes), arising, e.g., because of disorder.

The two-band Hamiltonian Eq.~(\ref{ham}) is reminiscent of the well-known Jaynes-Cummings model, however it allows for the dynamics of electron and hole pairs. The terms $J_j^e$ and $J_j^h$ represent the quantum mechanical tunnelling rates of the charged particles between neighbouring sites. In the absence of disorder ($J_j^e=J^e$, $\epsilon_j^e=\epsilon^e$ and $J_j^h=J^h$, $\epsilon_j^h=\epsilon^h$ for all $j$) and for $g_j\simeq g =0$, the first two terms in Eq.~(2) imply extended Bloch-type wave-functions for electrons and holes that can in principle lead to a finite conductivity even at zero temperature. However, disorder in semiconductors in 1D is known to invariably lead to wave-function localization and vanishing conductivity~\cite{Anderson}. Indeed this is shown in Fig. 4a for $g$=0 where the probability density of a few chosen electronic states is shown to be exponentially localized for a situation where we have chosen disordered hopping rates $J_j^{e,h}=J+\delta J_j^{e,h}$ and energy shifts $\epsilon_j^{e,h}=\epsilon +\delta \epsilon_j^{e,h}$, with $\delta J_j^{e,h}$ and $\delta \epsilon_j^{e,h}$ sets of random values (see Supplementary Section VI). 

Upon coupling the system to the vacuum electromagnetic field, it is expected that polaritonic states delocalized throughout the mode volume should appear \cite{FeistPRL2015,SchachenmayerPRL2015}.  This is shown in Fig. 4b (black curve) for $g=J$, where, remarkably, the electronic component of the polaritonic density is shown to be delocalized despite the strong disorder.  This value for $g$ was chosen so that the collective coupling $g\sqrt M$ is larger than all other coherent and incoherent rates, allowing for reaching the strong coupling regime (see Supplementary Section VI). Surprisingly, in addition to the polaritonic state we find that under conditions of strong coupling also other electronic states in the conduction band of the model are modified by the hybridization with the vacuum field and can become delocalized over the same length of the sample as shown in Fig. 4b for a few examples. This should have a considerable impact on conductivity, which we investigate next.

To probe the dynamics, we solve the master equation for the full density matrix of the system (see Supplementary Section VI) in the limit of low charge density. In order to mimic experiments, the master equation includes dissipative processes such as spontaneous decay of the plasmon mode and dephasing of the electron and hole wave-functions arising, e.g., from local molecular vibrations. Electrons are incoherently pumped into the system at a rate $\gamma_P$ , similar to the effects of an applied voltage. Fig. 4c shows the equivalent of the I-V curves derived from this model for $M=30$ sites and both $g=0$ and $g=J$. As can be seen, the current increases significantly in the presence of strong coupling. In the Supplementary Section VI we show that in fact significant enhancement is obtained only in the strong coupling regime, where the electronic wave-functions hybridized with light are delocalized.

Despite its simplicity, the model clearly shows that the origin of the enhanced mobility under strong coupling can be explained by the extended coherence of the states induced by the hybridization with the vacuum field. This occurs over lengths that correspond to the mode volume (on the scale of the wavelength of the resonance) in contrast with the normal situation where the carrier is mostly confined to the molecular (or a few molecule) scale, in particular at room temperature. 


It is important to note that, like for all conductors, disorder can strongly hinder charge transport whatever the transport mechanism. This is clearly shown in the case of PDI2EH-Br$_2$ which has a similar molecular structure to PDI2EH-CN$_2$, but which has much lower mobility~\cite{28} due to disorder. We found extremely weak resonant enhancement, if any, despite the strong coupling (see Supplementary Fig. S2). The P(NDI2OD-T2) polymer presents an intermediate case where despite the lower conductivity, the latter is enhanced by strong coupling. This is also confirmed by the model in the presence of large disorder, where, while the relative improvement of conductivity increases with disorder strength for a fixed light coupling $g$ over the case $g=0$, the whole conductivity tends to vanish.

The enhanced currents as well as the improved mobilities reported here demonstrate the potential of strong coupling for boosting conductivity in organic semiconductors. It is worth noting that at the same time the semiconducting properties are preserved. While we used surface plasmon resonances to couple to the organic semiconductors, metal micro-cavities or dielectric resonators could also be used provided that the mode volumes are sufficiently small that the strong coupling condition can be met. In particular metallic Fabry-Perot cavities lead to even larger Rabi splittings than plasmonic arrays which is important to increase the fraction of coupled oscillators in the sample, as we have shown elsewhere~\cite{18}.  The challenge, we have found, is to ensure that there is no leak via the top mirror when measuring the current parallel to the cavity. 

Conductivity of strongly coupled systems can be implemented with ease in a variety of designs and configurations, including in plastics, depending on the application.  Finally, our results show the importance of engineering the electromagnetic environment of a material not just for optical purposes but also for modifying its properties through the formation of hybrid light-matter states. This opens new prospects for material science as well as device technology. \\

{\it Note added}: Since submission of this work, two theoretical works addressed the question of suppression of backscattering of conduction electrons for a two-dimensional electron gas off-resonantly coupled to a high-frequency classical electromagnetic field~\cite{Kibis2014, Morina2015}.\\

{\it Acknowledgments}: This work was supported in part by USIAS, the ERC through the projects 
Plasmonics (227557), Suprafunction (257305), and Coldsim (307688), the 
International Center for Frontier Research in Chemistry (icFRC, 
Strasbourg), the ANR Equipex Union (ANR-10-EQPX-52-01), the Labex NIE 
projects (ANR-11-LABX-0058 NIE) and CSC (ANR-10-LABX-0026 CSC) within 
the Investissement d'Avenir program ANR-10-IDEX-0002-02, RYSQ, as well 
as the NSF (PIF-1211914 and PFC-1125844) and the Austrian Science Fund (FWF) via the project P24968-N27. Computations utilized the 
Janus supercomputer, supported by NSF (CNS-0821794), NCAR, and CU 
Boulder/Denver. \\

$^\dagger$These authors contributed equally to this work.


\begin{figure*}[t]
\centerline{\includegraphics[width = 0.7\textwidth]{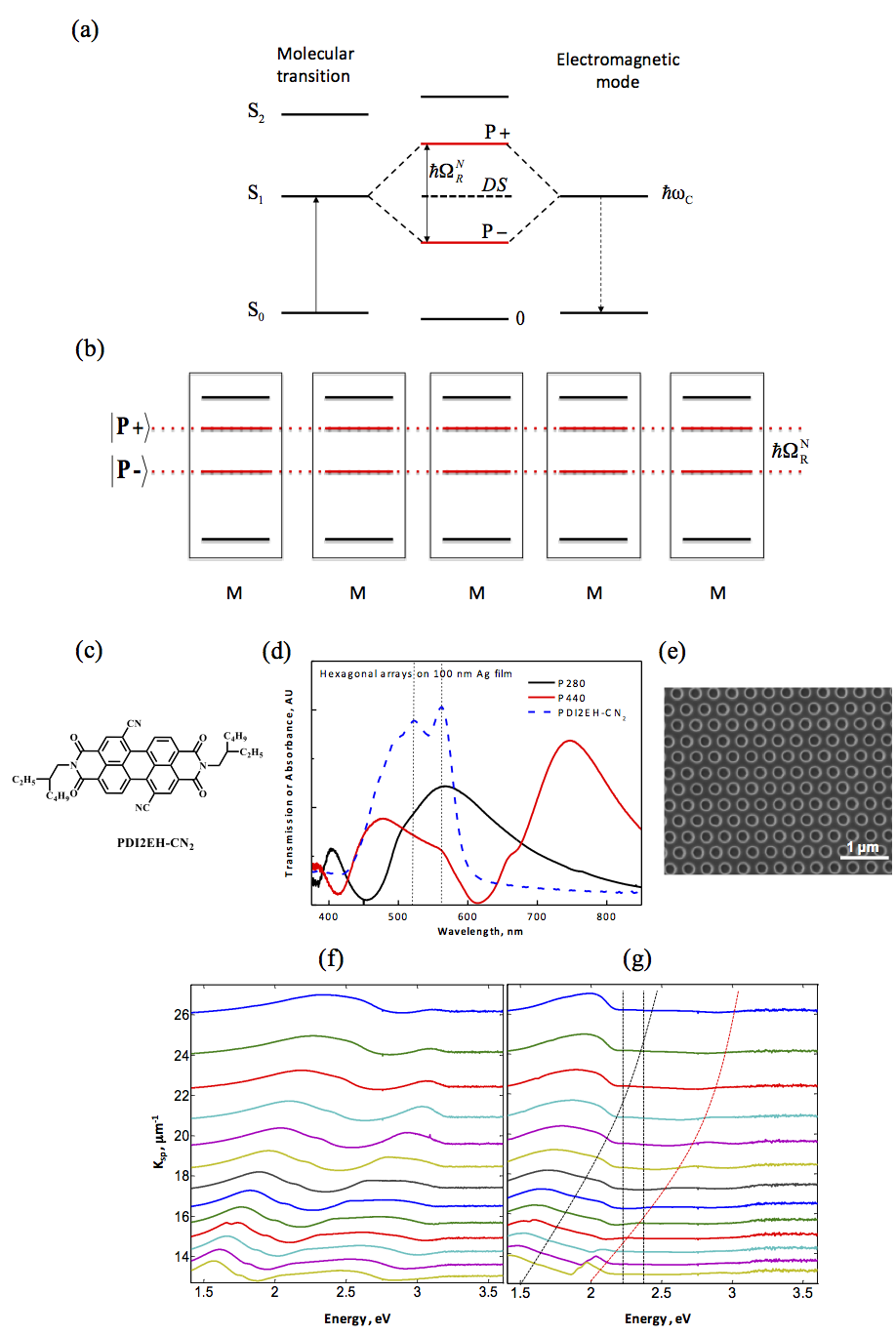}}
\caption{{\bf Light-matter strong coupling with organic semiconductors}. (a) Illustration of the strong coupling between a molecular electronic transition (S0 to S1) and an optical cavity mode of the same energy ($\hbar\omega_{\rm C}$). Two new hybrid light matter states P+ and P- are formed separated by the Rabi splitting $\hbar \Omega_{\rm R}^{\rm N}$. The energy level of the so-called dark states (DS) is also indicated. (b) Illustration of the collective nature P+ and P- forming a coherent state involving many molecules or molecular units (M). (c) Molecular structure of PDI2EH-CN$_2$. (d) Absorbance spectra of thin film of PDI2EH-CN$_2$ (dashed blue) and example of resonances of two hole arrays whose resonance matches the center of the molecular absorbance. (e) Electron microscope image of a hexagonal array milled in a Al metal film which provides the surface plasmon resonance that is strongly coupled to the molecule. Dispersion of the surface plasmon modes in the absence (f) and in the presence (g) of PDI2EH-CN$_2$ as a function of period P between 240 and 480 nm.  In the latter, the dashed lines are guides to the eye, the vertical ones indicating the position of the PDI2EH-CN$_2$ absorption while the curved ones indicate the shift in the plasmon resonances.} 
\label{Fig1}
\end{figure*}

\begin{figure*}[t]
\centerline{\includegraphics[width = 0.45\textwidth]{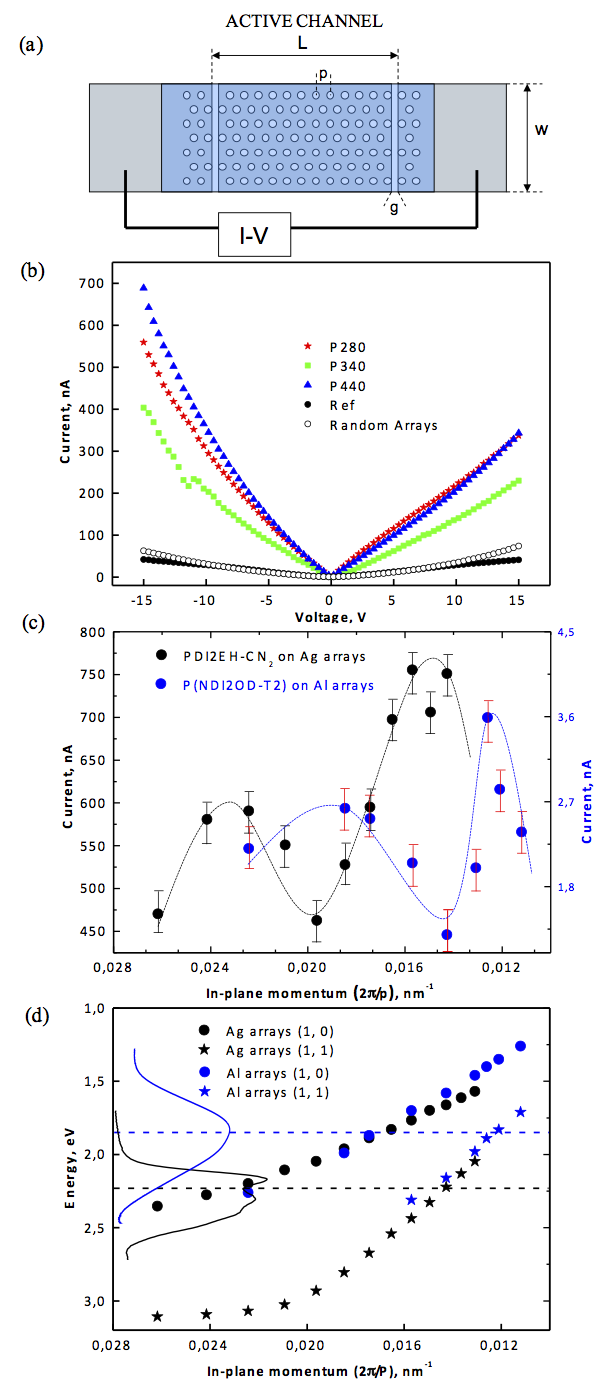}}
\caption{ {\bf Conductivity measurements under strong coupling on Ag and Al arrays}. (a) Illustration of the configuration used to measure conductivity using surface plasmon resonances generated by the hexagonal array milled in a 100 nm thick Ag film, 50 $\mu$m wide, deposited on a glass substrate. The distance between the electrodes is 50 $\mu$m. PDI2EH-CN$_2$ was spin-coated or evaporated forming a 100 nm thick layer on the array. The hole diameter is half the period. The gap between the array and the electrodes is 200 nm. (b) I-V curves as a function of the hexagonal array at selected periods for the configuration of Fig. 2a. (c) Current measured at $\pm$15V, and averaged over three separate runs, plotted as a function of period $2\pi / P$ for PDI2EH-CN$_2$ (black circle) and P-NDI2OD-T$_2$ (blue circle) showing two distinct resonances. (d) The two resonances seen in the current of Fig. 2c correspond to the intersection of the molecular absorbance and the array plasmon resonance, (1,0) and (1,1) modes of Ag (black) and Al (blue) arrays on the substrate side. At these intersection is where the strong coupling occurs. } 
\label{Fig2}
\end{figure*}

\begin{figure*}[t]
\centerline{\includegraphics[width = 0.8\textwidth]{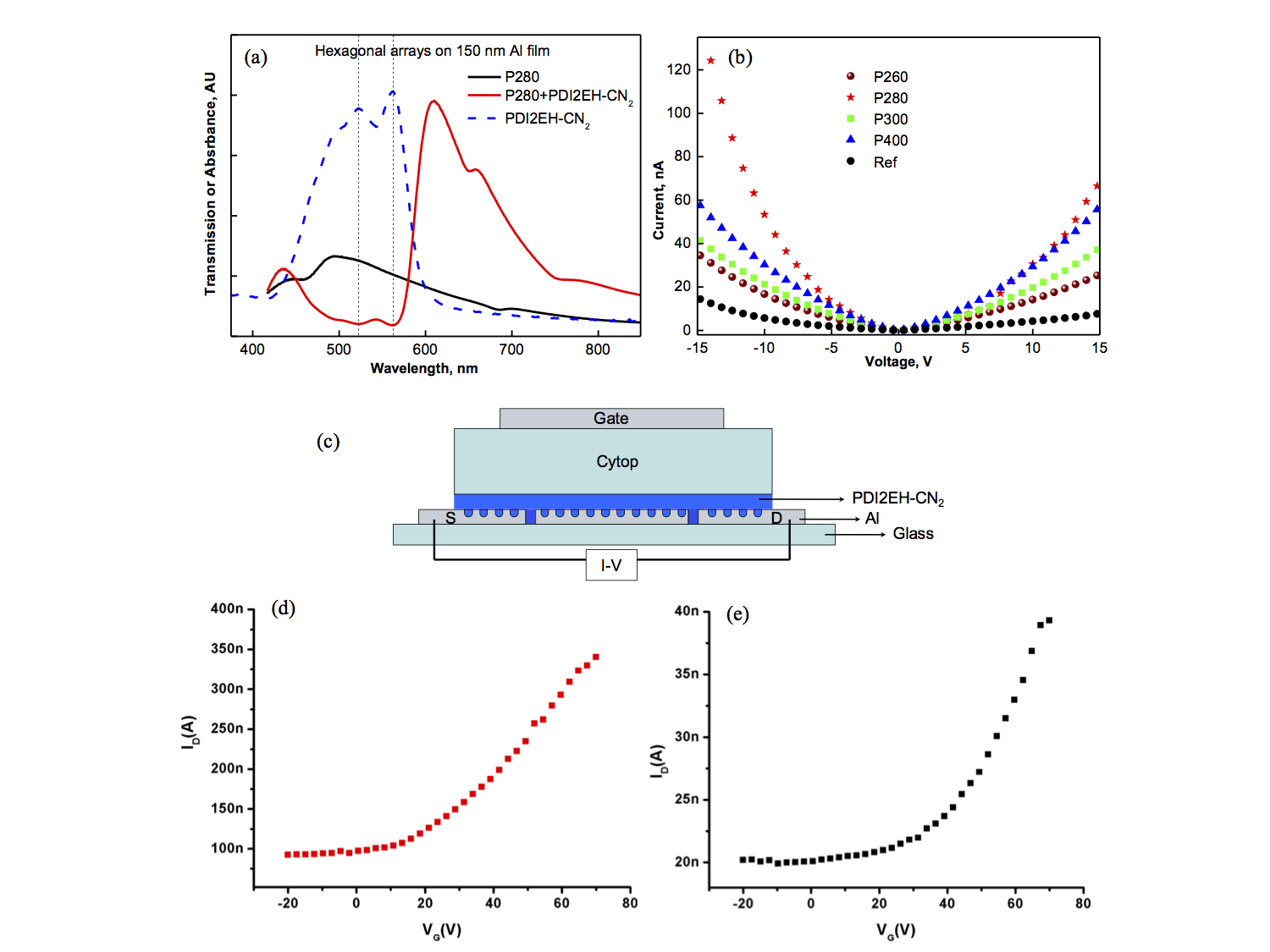}}
\caption{ {\bf Conductivity under strong coupling on Al arrays and Gating}. 
(a) Spectra of Al hole array with and without PDI2EH-CN$_2$ as shown; (b) I-V curves for selected periods versus reference in the same geometry as in Fig. 2a. (c) Three terminal gating device geometry. It is composed of:  Al (40 nm) acting as a top gate; $\sim 600$ nm of Cytop, a polymer dielectric acting as a gate insulator ($\varepsilon_r = 2.1$); 100 nm of PDI2EH-CN$_2$ as the semiconductor layer; Al electrodes and nanohole array structure. [W = 50 $\mu$m, L = 50 $\mu$m]. Note that the surface plasmon modes is only slightly modified by the presence of the dielectric as most of the evanescent plasmonic field is confined within the organic semiconductor layer. Typical transfer characteristics ($I_D-V_G$) of top-gate bottom-contact FETs realized on the structure proposed in Figure 3c. In particular the graph shows the difference between a gated structure with hole arrays of periodicity $P = 280$ nm (d) vs. a reference (e) structure with the same geometry but without nanohole arrays. The ratio of the mobilities is $\mu_{\rm FET [P = 280 nm]} / \mu _{\rm FET[reference]} \sim10$ which corresponds essentially to the ratio of the current measured in the I-V curves. } 
\label{Fig3}
\end{figure*}

\begin{figure*}[t]
\centerline{\includegraphics[width =0.9 \textwidth]{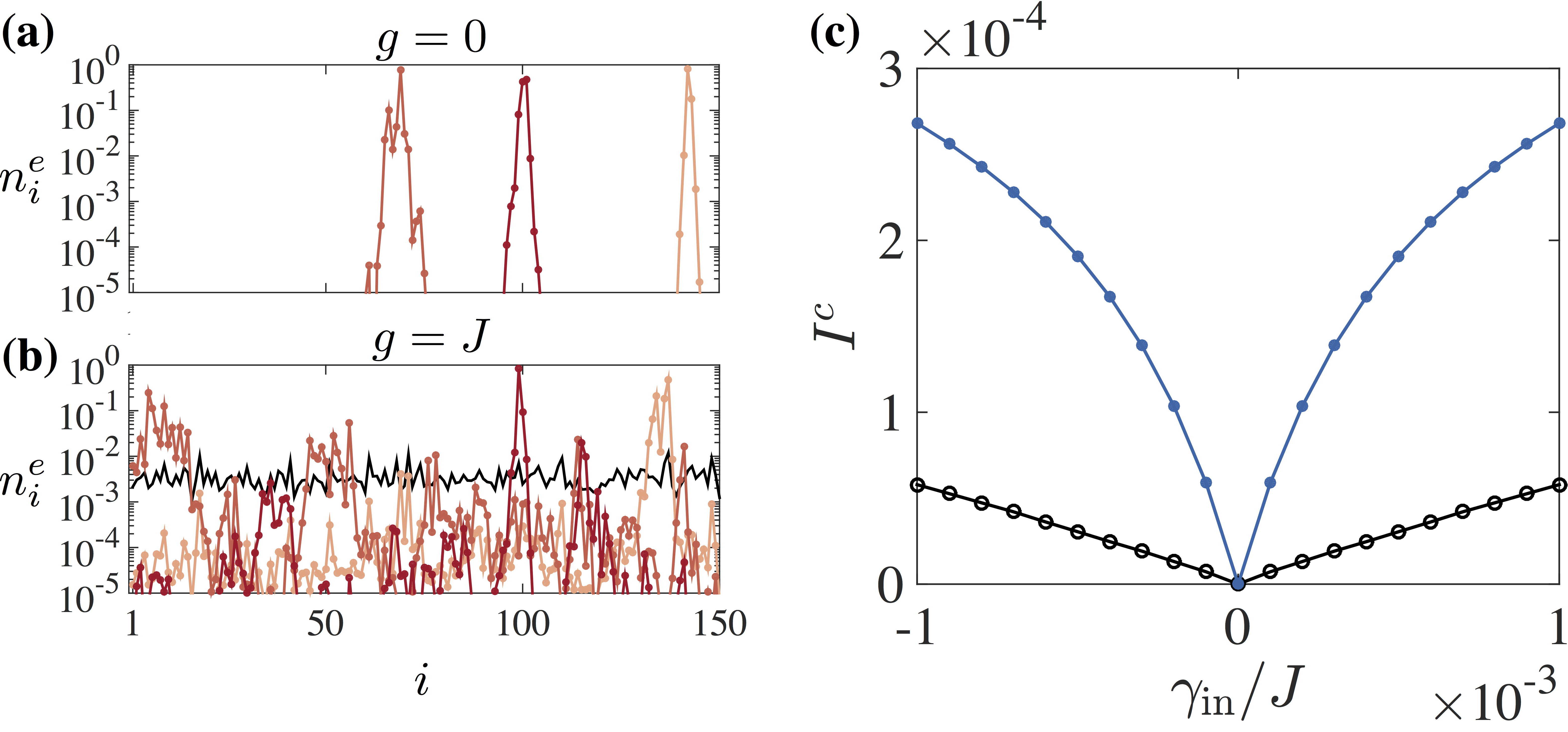}}
\caption{(color online). {\bf Theoretical model}. (a) Electron density $n_i^e$ as a function of the site (molecule) position $i$, for a chain of $M=150$ molecules, in the presence of disorder (see text) and in the absence of plasmon - molecule coupling $g=0$. Because of disorder, $n_i^e$ is localized at individual sites, as expected. (b) $n_i^e$ {\it vs.} $i$ for a finite plasmon - molecule coupling  $g=J$ in the strong coupling regime. Suprisingly, electronic wave-functions (and thus $n_i^e$) become delocalized over the length of the sample. Black  line: lower polariton. Other continuous lines: $n_i^e$ corresponding to those of panel (a). (c) Electronic current $I^{c}$ {\it vs.} rate of electron pumping $\gamma_{\rm in}$, analogous to $I-V$ curves for a chain with $M=30$ molecules. Black empty dots and blue full dots corresponds to the cases $g=0$ and $g=J$, respectively.} 
\label{Fig4}
\end{figure*}

\newpage 

\begin{center}
{\bf Supplementary Material for: Conductivity in Organic Semiconductors Hybridized with the Vacuum Field}
\end{center}

\section{I. Materials}

PDI2EH-CN$_2$ was synthesized from the PDI2EH-Br$_2$ (Suna Tech Inc) with all the precursors for the synthesis purchased from Sigma-Aldrich and following the published procedure \cite{JonesJACS2007}. To a 250 mL round bottom flask was added (1.9 mmol) 1.47 g of PDI2EH-Br$_2$ and 3.1 g (34.6mmol) Cu(I)CN followed by the addition of 75 mL dry DMF and the mixture was heated to 150$^o$C under nitrogen atmosphere overnight. The product was concentrated and continuously Soxhlet extracted with CHCl$_3$ for 2 days. The extracted mixture was rotavapped to dryness and the crude product chromatographed on silica, with chloroform/ethylacetate, slowly from 1\% to 2\%. The product was then precipitated 3 times from chloroform with methanol give 0.82 g (1.23 mmol) of PDI2EH-CN$_2$ in 65\% yield. $^1$H NMR (CDCl$_3$): $\delta$ 9.69-9.65 (d, 2H), 8.95 (s, 2H), 8.92- 8.88 (d, 2H), 4.22-4.16 (t, 4H), 1.89-0.89 (m, 30H).

\begin{figure}
\includegraphics[width=0.95\columnwidth]{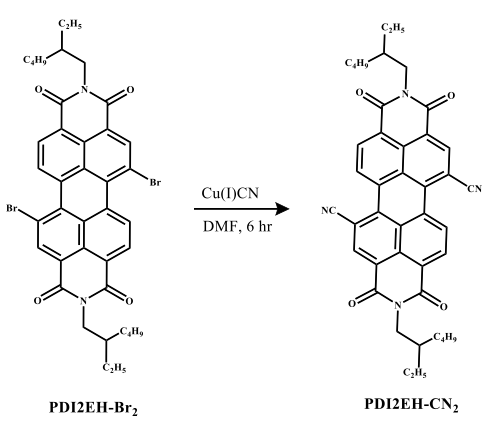}	
\caption{Scheme of PDI2EH-CN$_2$ synthesis}
\label{S1}
\end{figure}

\section{II. CURRENT-VOLTAGE MEASUREMENTS AND ELECTRICAL CHARACTERIZATION} 

\subsection{Two- and three-terminal devices including nanohole arrays}

All the electrodes were realized on a glass substrate BK7 (25 x 25 mm) after standard cleaning procedures with sonication in Alconox solution (0.8$\%$ solution in milliQ water), then rinsed with water and sonicated for 1 hour in (spectroscopically pure) ethanol. The glass substrates are then dried in a oven overnight. Ag and Al electrodes (100 nm) were fabricated using an electron beam evaporator (Plassys ME 300) at optimized working pressure ($\sim 10^{-6}$ mbar) and deposition rates ($\sim 0.4$ nm/s). Plasmonic hole-arrays are generated by NPVE software programme and milled using Carl Zeiss-Cross Beam Origa-FIB system. 0.5 wt$\%$ of PDI2EH-CN$_2$ or 0.5 wt$\%$ of PDI2EH-CN$_2$ or 1 wt$\%$ P(NDI2OD-T2) solutions were prepared by dissolving the molecules in spectroscopic grade anhydrous chloroform in an inert atmosphere and spincoated onto the electrodes at 750 rpm and 4000 rpm, respectively, to achieve 100-nm thick films. 

Regarding the fabrication of field-effect transistor devices starting from the electrode pairs bearing nanohole arrays, the two-terminal structures described above were completed by spin-coating of a commercial fluoropolymer (Cytop 809-M) at 2500 rpm leading to 700-nm thick dielectric layers. The dielectric layer, processed from an orthogonal solvent which prevents the underlying PDI film to dissolve, would act as the gate insulator for the three-terminal device whose fabrication was completed by thermal evaporation ($P\sim 10^{-6}$ mbar, evaporation rate $\sim 1$ nm/s) of a 60 nm thick Al gate at the top (see Main Text Fig. 3c). Mobilities and threshold voltage values were extracted as detailed later in this session. 

The electrical characterization of both two- and three-terminal devices was carried out by means of a Cascade Microtech MPS probe station equipped with micro-positioners to contact the electrode pads. The $I-V$ (2-terminal device) and $I_D-V_G$ (3-terminal device, $I_D$ drain current, $V_G$ gate voltage) characteristics are recorded through a Keithley 2636A source-meter interfaced with {\sc LabTracer} 2.0 software. All the experiments were repeated more than 6 times and the current enhancement is plotted based on averaged and different runs as shown in Main Text Fig. 2.

\subsection{Extraction of the contact resistances}

The current-voltage measurements were carried out with a Keithley 2635 on two-terminal devices with thin films of the PDIEH-CN$_2$. The measurements were performed in a glovebox under N$_2$ environment in order to avoid contamination from oxygen and water molecules which could induce (i) intentional doping or degradation of the semiconductor and/or (ii) oxidation of the injecting metal electrodes and therefore alter the outcome of the measurements. 

The total resistance vs. the electrode distance $L$ is plotted in Fig. S2. With an electrode width $W$, three different runs were averaged to obtain the correspondent $R \times W$ value at a given $L$. The contact resistance, $R_{\rm contact}$ is the intercept of the fitting line with the $Y$ axis and amounted to 9 $\Omega\cdot$cm and 0.3 $\Omega\cdot$cm for Al and Ag, respectively. The reported values are consistent with previous findings on very similar systems \cite{Boudinet}.

\begin{figure*}[tb]
		\includegraphics[width=0.95\textwidth]{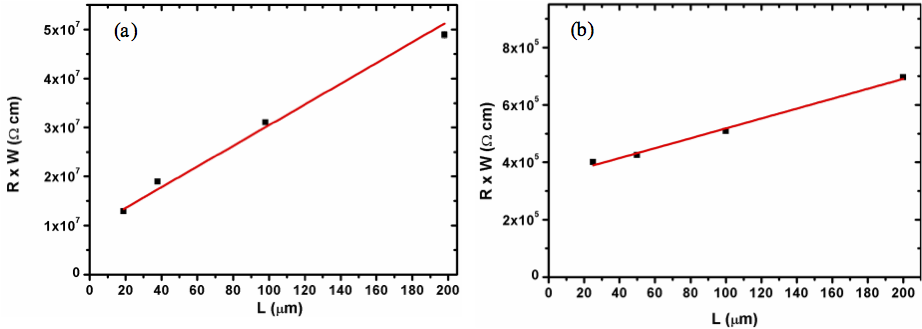}
\caption{Plot of $R \times W$ ($\Omega\cdot$cm) {\it vs.} electrode gap for Al (a) and Ag (b) measured for 100 nm-thick films of PDI2EH-CN$_2$. The electrode width $W$ and thickness $t$ equal 50 $\mu$m and 100 nm, respectively, for all samples.}
	\label{2S}
\end{figure*}

\subsection{Classical electrical characterization of the materials encompassed in this study}

Organic thin-film transistors were prepared starting from n$^{++}$-Si/SiO$_2$ substrates exposing pre-patterned interdigitated gold source and drain electrodes (purchased by the IPMS Fraunhofer Institute). After an accurate cleaning via sonication in acetone and isopropanol, the substrates underwent a hexamethyldisilazane (HMDS, purchased by Sigma-Aldrich) treatment which helps improving the surface quality and to passivate possible electron trap states coming from the hydroxyl groups of the silanols (Si-OH) present at the silicon oxide surface. Afterwards, solutions of the different materials were spincast onto the substrates which underwent electrical characterization (performed in an inert environment) by means of an electrometer, Keithley 2636A, interfaced by {\sc LabTracer} software.

For the sake of consistency, the three materials, namely  PDI2EH-CN$_2$,PDI2EH-Br$_2$ and P(NDI2OD-T2), were deposited on the n$^{++}$-Si/SiO$_2$ substrates from spin-coating in the exactly same experimental conditions as described in the previous section. 

The following equations correlate the saturated drain current of a n-type thin-film transistor with the applied bias voltages to the gate and the drain electrode respectively while the source electrode is grounded. 

In particular considering the following equation
\begin{eqnarray}
I_D=\frac{1}{2}\mu_{\rm SAT,n}C_i\frac{W}{L}\left(V_G-V_{\rm TH}\right)^2
\label{eS1}
\end{eqnarray}

which holds in the saturation regime, where $V_D\leq V_G-V_{\rm TH}$. Here, $I_D$ is the drain current, $V_G$ is the gate voltage, $V_D$ is the drain voltage, $L$ is the channel length, and $W$ the channel width, and $C_i = 1.5\cdot 10^{-8}$ F$\cdot$cm$^{-2}$ is the capacitance of the gate dielectric per unit area. The field-effect mobility for electrons in the saturation regime can be extracted from 
\begin{eqnarray}
\mu_{\rm SAT,n}=\frac{2}{C_i}\frac{L}{ W}\left(\frac{\partial \sqrt{I_D}}{\partial V_G}\right)^2
\label{eS2}
\end{eqnarray}
The threshold voltage ($V_{\rm TH}$) values can be extracted by plotting $\sqrt{I_D}$ vs. $V_G$ curve and then extracting the intercept with the voltage axis of the line which fits the linear part of the curve.

The major electrical parameters measured are gathered in Table I below.
\begin{table*}[t]
\centering
\begin{tabular}{| l |}
 \hline
PDI2EH-CN$_2$ : \\
\hline
$\mu_{\rm SAT}  = 2.2\times 10^{-3}$ cm$^2$V$^{-1}$s$^{-1}$ / $V_{\rm TH} = +1$ V  \\
\hline
PDI2EH-Br$_2$ : \\
\hline
$\mu_{\rm SAT}  = 3.4\times 10^{-6}$ cm$^2$V$^{-1}$s$^{-1}$ / $V_{\rm TH} = +33$ V  \\
\hline
P(NDI2OD-T2) : \\
\hline
$\mu_{\rm SAT}  = 7.0\times 10^{-5}$ cm$^2$V$^{-1}$s$^{-1}$ / $V_{\rm TH} = -19$ V  \\
\hline
\end{tabular}
\caption{Measured electrical parameters.}
\label{table1}
\end{table*}

\begin{figure*}[tb]
		\includegraphics[width=0.95\textwidth]{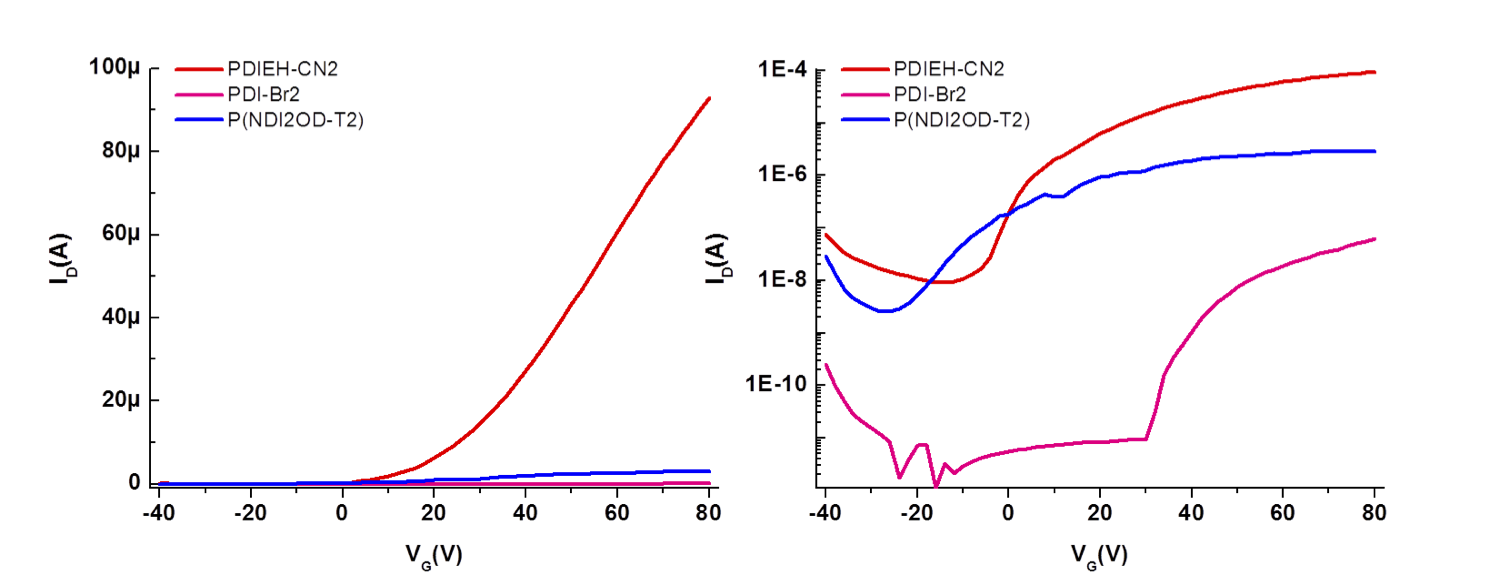}
\caption{Comparative transfer curves ($I_D - V_G$) plots in linear (left) and log scale (right) acquired in the saturation regime (at a saturation drain voltage $V_{D} = 60$ V) measured for the three materials with fixed channel dimensions ($W = 10000~\mu$m, $L = 20~\mu$m).}
	\label{3S}
\end{figure*}

\begin{figure*}[tb]
		\includegraphics[width=0.95\textwidth]{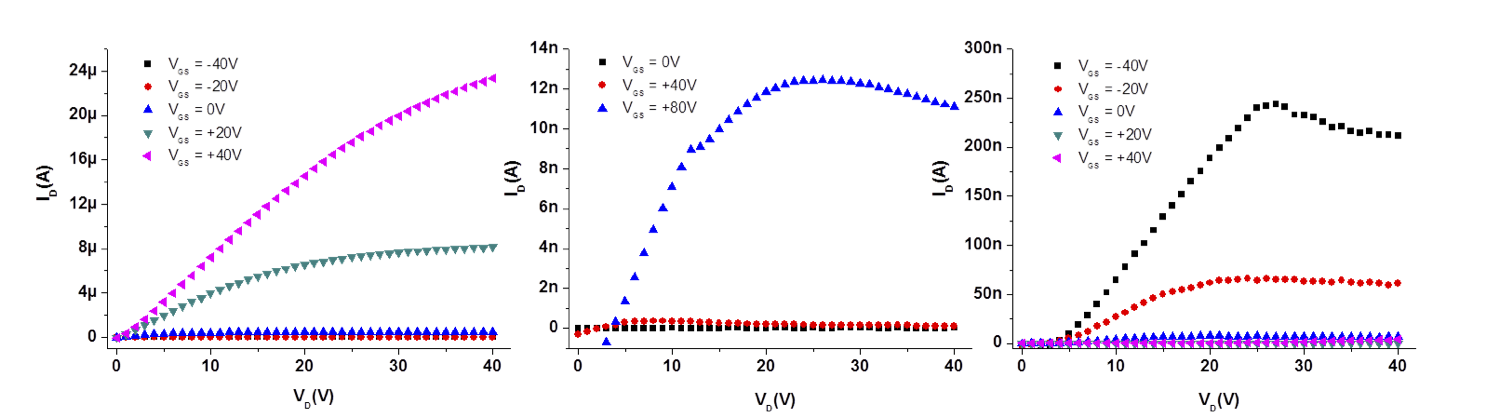}
\caption{ Output characteristics ($I_D - V_D$) of PDI2EH-CN$_2$ (left), PDI2EH-Br$_2$ (center) and  P(NDI2OD-T2) transistors with fixed channel dimensions ($W = 10000~\mu$m, $L = 20~\mu$m).}
	\label{4S}
\end{figure*}

\section{III. Data for PDI2EH-Br$_2$}

\begin{figure*}[tb]
		\includegraphics[width=0.8\textwidth]{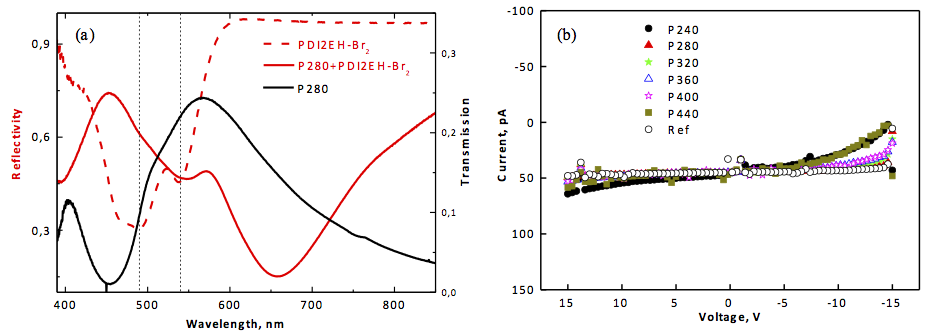}
\caption{(a) Reflectance spectra of thin film of PDI2EH-Br$_2$ (dashed red) and that of combined hole array plus PDI2EH-Br$_2$ (red) together with transmission of the corresponding hole array (black); (b) the I-V curves as a function of the hexagonal array at selected periods for the configuration of Fig. 2a measured for 100 nm-thick films of PDI2EH-Br$_2$. Notice that the pA currents measured are at the limit of current resolution.}
	\label{S5}
\end{figure*}

The data are gathered in Fig. \ref{S5}.

\section{IV. CHARACTERIZATION OF THE ENERGY LEVELS OF METALS AND THE THREE SEMICONDUCTORS}

\subsection{Ambient photoelectron spectroscopy measurements}

Ambient photoelectron spectroscopy measurements were performed by sampling in each measurement an area of about 4 mm$^2$ (beam size) by using a Photoelectron Yield Spectrometer operating in Ambient environment (PYSA), Model AC-2 from Riken Keike Co., Ltd. The semiconductor films were prepared by spin-coating onto a conductive Au/glass substrate which would form a film exceeding 10 nm therefore sufficient to ensure that only the top material was probed. 

The metal Al and Ag films on glass substrates were prepared as previously described and measured over time to determine whether the exposure to air would affect their respective work function. In all metal films employed as electrodes in the electrical characterization measurements, the exposure to air of all samples is minimized and it amounts to some minutes in total which includes the transfer time from the evaporating chamber into the FIB chamber (UHV at $10^{-6}$ mbar) and to the latter into the glovebox (nitrogen). 

The ionization energy of P(NDI2OD-T2) amounted to ($5.55 \pm 0.05$) eV in perfect accordance with previous measurements \cite{YanNature2009}. By considering an optical band gap of $1.55$ eV as reported in literature \cite{MoriScience2014}, it was possible to estimate the energy of the LUMO level of P(NDI2OD-T2) as ($4.00 \pm 0.05$) eV.

The ionization energy measured on Al and Ag films corresponds to their absolute work function. For Al films the work function was found to be constant with time and to amount to ($4.2 \pm 0.05$) eV.  For Ag films, the variation of the work function over a period of time up to 4 days was monitored. The work function changed from 4.72 eV (fresh sample) up to 4.81 eV (3 days 14h) which, including the experimental error (normally ca. 50 meV), gives a net variation of 50 meV approximately. The Ag work function was found to be 4.75 eV after some minutes of exposure to air which most probably corresponds to the value the Ag work function during the device experiments.

\begin{table*}[t]
\centering
\begin{tabular}{| l | c | r |}
 \hline
  $E^1_p$   & $E^2_p$ & LUMO$^a$   \\
 \hline
  -0.716 V & -1.023 V & 4.08 eV \\
 \hline
\end{tabular}
\caption{PDI2EH-CN$_2$ reduction potentials as measured in Fig. \ref{S7}. $^a$: $E^{\rm LUMO}= [E^{\rm red}-E_{\frac{1}{2}(Fc/Fc^+)}]$.}
\label{table2}
\end{table*}

\begin{figure*}[tb]
		\includegraphics[width=0.8\textwidth]{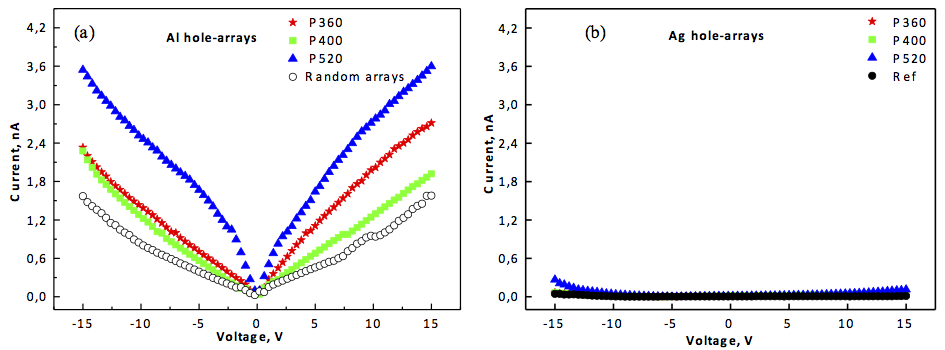}
\caption{$I-V$ curves as a function of hexagonal hole-arrays of (a) Al and (b) Ag at selected periods for the configuration shown in Fig. 2a of the main text, with a $100$ nm-thick film of P(NDI2OD-T2). }
	\label{S6}
\end{figure*}

\subsection{Cyclic voltammetry measurements}

The ionization energy of both PDI derivatives was found to fall outside the detection of the PYSA which only allows to reliably measure occupied energy states up to approximately 6 eV. However, the optical band-gap (Eg$_{\rm opt}$) could be estimated from the most intense peaks of the absorbance spectra (in thin films) which correspond to the 0-0 transitions which were placed at $\lambda = 540$ nm  and $\lambda = 560$ nm for PDI2EH-Br$_2$ and PDI2EH-CN$_2$, respectively. The values of Eg$_{\rm opt}$ were therefore of ca. 2.3 eV for the former and ca. 2.1 eV for the latter molecule. For PDI2EH-Br2 this value corresponds to previous reported values \cite{JonesJACS2007} which would then place the LUMO at 4.0 eV. 

\begin{figure*}[tb]
		\includegraphics[width=0.6\textwidth]{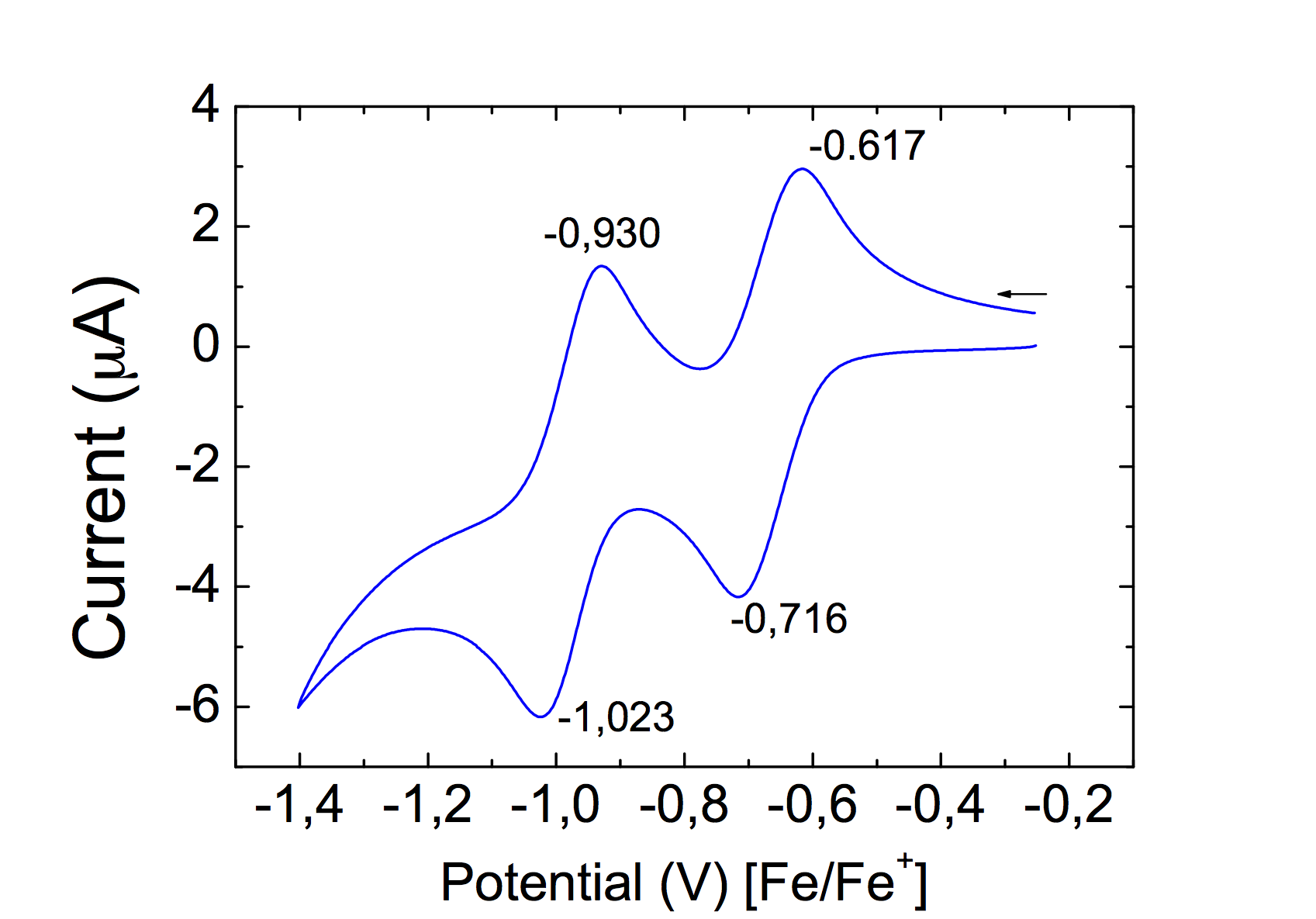}
\caption{Cyclic voltammetry measurements of a PDI derivative bearing -CN groups and lateral branched alkyl chains. }
	\label{S7}
\end{figure*}

In order to determine the LUMO of PDI2EH-CN$_2$ cyclic voltammetry measurement were performed on a PDI derivative bearing -CN groups and lateral branched alkyl chains. Cyclic voltammetry was performed using an Autolab (Eco Chemie B.V.) potentiostat interfaced to a PC with {\sc Autolab} software. A three-electrode configuration contained in a non-divided cell consisting of a platinum disc (d = 1 mm) as working electrode, a platinum wire as counter-electrode and a Ag/AgCl as reference electrode was used. Measurements were carried out in dichloromethane containing 0.1 M of Bu$_4$NPF$_6$ and using a scan rate of dE/dt = 0.01 V$\cdot$s$^{-1}$. The data are given in reference to the ferrocene redox couple (Fc/Fc$^+$), which was used as external standard. The cyclic voltammogram of the above-mentioned PDI derivative is shown in Fig. \ref{S7} and its reduction potentials are gathered in Table II. Hence, PDI2EH-Br$_2$ and PDI2EH-CN$_2$ the energy level of the LUMO sits at a nearly identical (within experimental error) value of 4.0 eV.


\section{V. RANDOM ARRAYS AND CROSS SECTION ANALYSIS} 

\begin{figure*}[tb]
		\includegraphics[width=0.8\textwidth]{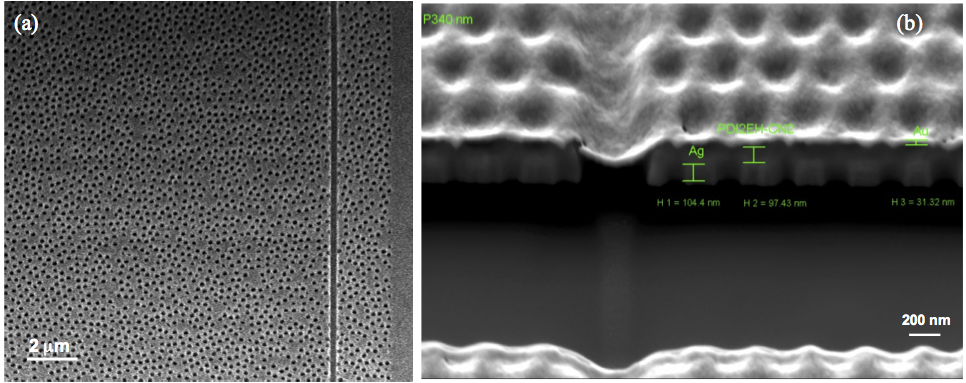}
\caption{(a) SEM image of the edge of a random array milled on 100 nm Ag electrode; (b) Cross section analysis of 100 nm-thick films of PDI2EH-CN$_2$ on 100 nm Ag hole-array electrode. Please note that 30 nm Au film was sputtered on the PDI2EH-CN$_2$ semiconductor to avoid charging effect while SEM measurement. }
	\label{S8}
\end{figure*}

Random array structures were used as another reference to make sure that the observed effect is not due to conformality of molecules on hole-arrays on the metal film. Random arrays were generated from a {\sc Matlab} code comparable to the code used for hexagonal arrays and then milled into the electrodes on $55\times 80~\mu$m$^2$ area. Then, as for the other electrodes, two cuts $50~\mu$m apart were milled into it. We tried different densities of random holes and one example (detail of a bigger array) with a high density is shown on Fig. \ref{S8}(a). The conductivity experiments revealed that the currents measured over the electrode pairs comprising a random non-resonant array are comparable or very close to those measured over samples bearing no nanoholes (flat metal film in between cuts). 

For cross section analysis 30 nm Au was sputtered on to the best performing electrodes with the hole-arrays and a thin layer of semiconductor molecules on top. Sample was introduced to the FIB chamber, cut through the array electrode until it reaches the glass substrate and SEM images were acquired. Figure \ref{S8}(b) is an example showing a sandwich layer of organic semiconductor molecule (~100 nm, as expected) between 100 nm thin metallic film of arrays and 30 nm sputtered Au film. It is clear from the SEM measurement that the spin coating of organic molecule on hole-arrays are conformal and are homogeneously spread to have a nice coverage. Further, it should be noted that the topography features are such that they do not disrupt the structure of the semiconductor as these features are over scales of hundreds of nm, at least the periodicity of hole arrays, which is way beyond the scale at which one would observe spatial confinement effects on the semiconductor.

\section{VI. Dynamics in the theoretical model}

We treat our model as an open quantum system with dynamics governed by the following master equation for the reduced density matrix $\rho$ ($\hbar$=1) 

\begin{eqnarray}
\frac{d}{dt}  \rho = -\text{i} [ H,  \rho] + \mathcal{L}^{{\rm in},e} ( \rho)+ \mathcal{L}^{{\rm in},h} ( \rho)+ \mathcal{L}^{{\rm out},e}( \rho) \nonumber\\ +  \mathcal{L}^{{\rm out},h}( \rho) +   
\mathcal{L}^{{\rm dec}} ( \rho) + \sum_{i=1}^M \left[ \mathcal{L}^{{\rm deph},e,i} ( \rho) + \mathcal{L}^{{\rm deph},h,i} ( \rho) \right]. \label{meq}
\end{eqnarray}

The dissipative terms describe interactions with an environment and have the trace-preserving form

\begin{align}
	 \mathcal{L}^{\mu}( \rho)= 2  L_\mu  \rho  L_\mu^\dag -  L_\mu^\dag  L_\mu  \rho -  \rho  L_\mu^\dag  L_\mu.
\end{align}

\begin{figure*}[tb]
		\includegraphics[width=0.95\textwidth]{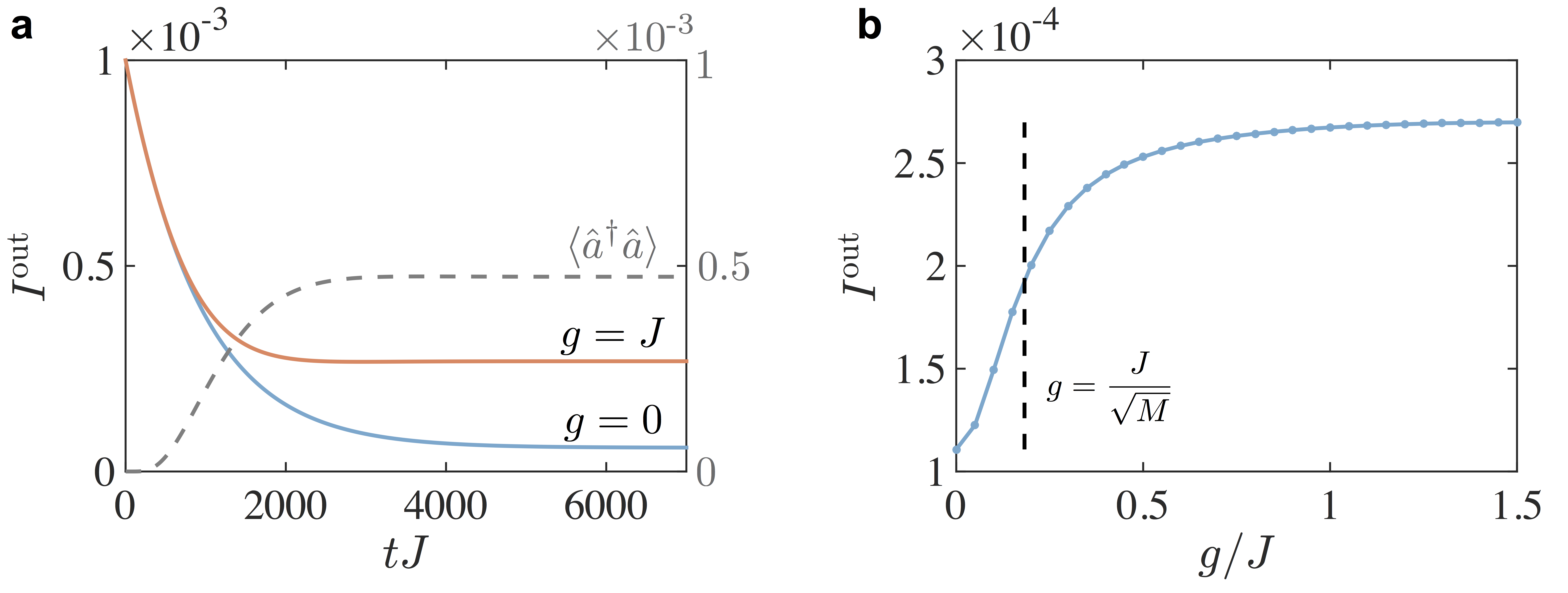}
\caption{(a) Time dependence of the electric current in the model, $I^{\rm out}$, for a pumping rate of $\gamma_P=10^{-3}J$ in a system with $M=30$ sites. In the presence of a plasmon coupling with strength $g=J$  (red line), the current is enhanced. The enhancement is correlated with an occupation of the plasmon mode (dashed grey line). (b) $I^{\rm out}$ for $\gamma_P=10^{-3}J$ in the steady-state ($tJ=5000$) as function of the coupling $g$. Enhancement starts to show up in the collective strong coupling regime, where $g\sqrt{M}\gtrsim J$ ($g\sqrt{M}={J}$ is marked by the dashed vertical line).}
	\label{fig:cur}
\end{figure*}

Here, $L_\mu$ are quantum jump operators for the particular dissipative processes, which have the form
\begin{align}
&L_{{\rm in},e}= \sqrt{\frac{\gamma_p}{2}} ( c_1^{e})^\dag,\quad
 L_{{\rm in},h}= \sqrt{\frac{\gamma_p}{2}}  c_1^h,\quad\nonumber\\
& L_{{\rm out},e}= \sqrt{\frac{\gamma_p}{2}}  c_M^e,\quad
 L_{{\rm out},h}= \sqrt{\frac{\gamma_p}{2}} ( c_M^h)^\dag,
\end{align}
and
\begin{align}
&&L_{{\rm dec}}= \sqrt{\frac{\kappa}{2}}  n_i^e, \nonumber  \\
&& L_{{\rm deph},e,i} = \sqrt{\frac{\gamma_{d}}{2}}  n_i^e, \nonumber  \\
 &&L_{{\rm deph},h,i} = \sqrt{\frac{\gamma_{d}}{2}}  n_i^h.
\end{align}
The terms in the first row mimic a situation of a voltage applied to our open quantum system. They describe the injection of negative electric charge from the environment to the left side ($i  = 1$) of the system at a pumping rate $\gamma_{\rm P}$ and its removal on the right side ($i = M$) at a similar rate. The terms in the second row are introduced to simulate the real situation of dissipation and noise in the system. Here, we consider a spontaneous decay of the plasmon mode with rate $\kappa$ as well as noise that destroys the quantum-wave nature of a given state. Dephasing terms arise for example from local vibrations that effectively "measure" the position of the particles (both electrons and holes) in the lattice and thus localize them on individual sites (molecules) at a rate $\gamma_{\rm d}$. In our simulations we solve the time-dependent problem governed by the master equation after starting in a vacuum state of the surface plasmons with no electron-hole pair excitation. As an observable, we calculate the charge current that leaves the chain on the right side. This quantity can be derived from a continuity equation for the total charge in the system, defined as $ C=\sum_{i=1}^M ( n_i^h -  n_i^e)$ (with $e\equiv 1$).  The time-evolution of $C$ is dictated by the master equation above and reads
\begin{align}
&\frac{d}{dt} \langle  C \rangle = \left\{\text{tr}[ C  \mathcal{L}^{{\rm in},e} ( \rho)]+ \text{tr}[ C \mathcal{L}^{{\rm in},h} ( \rho)] \right\}  \nonumber\\
& \left\{\text{tr}[ C \mathcal{L}^{{\rm out},e}( \rho)] 
 +  \text{tr}[\hat C \mathcal{L}^{{\rm out},h}( \rho) ] \right\}
\equiv I^{\rm in} +  I^{\rm out} 
\end{align}
Here, the Hamiltonian, the decay and dephasing terms of Eq.~\eqref{meq} do not contribute since $ H$, $ a$ and $ n_i^{e,h}$ commute with $ C$, i.e.~conserve the charge in the system. In the steady state $\frac{d}{dt} \langle  C \rangle=0$ and thus the terms $I^{\rm in}$ and $I^{\rm out}$ are identical in magnitude and  opposite in sign. We are interested in the charge currents that arise for particular pumping rates $\gamma_{\rm P}$ as shown in the main manuscript of the paper.\\

A typical lifetime of the plasmon mode is $t_{pl}=50\,\text{fs}$. In our units, time is quantified in inverse tunneling times $\hbar/J$, and thus for a typical $J=0.03\,\text{eV}$, we have $t_{pl}\approx 2 J$. We thus  use $\kappa=J/2$. Characteristic dephasing rates are on the order of  $100\,\text{fs}$ and we thus use $\gamma_{\rm d}=2J$. Disorder in the position of the particles leads to typical tunneling frequencies that vary by $20\%$ of the bare tunneling frequency. Thus in the Hamiltonian $H$ we choose $\delta J^e_i=0.2 \mathcal{N}(J)$, where $\mathcal{N}(J)$ is a random number drawn from a Gaussian distribution with standard deviation $J$. Our open system simulations are limited to relatively small system sizes of $M\sim 30$. Thus we choose the disorder in the energy offset $\epsilon^e_i=3 \mathcal{N}(J)$ to be large enough to make the exponential decay of the wave-functions relevant on the system length. In addition  we assume holes to be much less mobile than the electrons, and thus consider $\delta J^e_i=0.5 \mathcal{N}(J)$ and $\epsilon^h_i=5 \mathcal{N}(J)$.

After switching on the electrical pumping, the evolution converges to a steady state with a fixed charge current as shown in Fig.~\ref{fig:cur}a. We find that a finite plasmon coupling $g$ dramatically enhances the current in the system, which is a result of the delocalization of the hybrid states. In Fig.~\ref{fig:cur}a we also find that the enhancement is directly connected to the occupation number of the plasmon mode. Finally, considerable current enhancement is achieved in the collective strong coupling regime defined when $g\sqrt{M}$ is much larger than both decay $\kappa$ and dephasing $\gamma_{p,d}$ rates. Note that $g\sqrt{M}\gtrsim J$ becomes of the order of the largest energy scale of the uncoupled system, as shown in Fig.~\ref{fig:cur}b.

\end{document}